\documentclass{osa-article}

\journal{ome}
\usepackage{mathtools}	% fancy math
\usepackage[]{esdiff}	% used to get \diff and \diffp
\usepackage{chemformula}	% chemical formulae in text
\usepackage[separate-uncertainty = true, multi-part-units=single]{siunitx}	% support for units
\DeclareSIUnit\Molar{\textsc{m}}
\DeclareSIUnit\pulse{pulse}

\DeclareMathOperator*{\expintegral}{Ei}

\begin{document}

\title{Material gain concentration quenching in organic dye-doped polymer thin films}

\author{Florian Vogelbacher,\authormark{1,2,*}
Xue Zhou,\authormark{3}
Jinhua Huang,\authormark{3}
Mingzhu Li,\authormark{3}
Ke-Jian Jiang,\authormark{3}
Yanlin Song,\authormark{3}
Karl Unterrainer,\authormark{2}
and Rainer Hainberger\authormark{1}}

\address{
\authormark{1}AIT Austrian Institute of Technology GmbH, Center for Health \& Bioresources, Giefinggasse~4, 1210~Vienna, Austria\\
\authormark{2}TU Wien, Photonics Institute, Gusshausstr.\@ 27-29, 1040~Vienna, Austria\\
\authormark{3}Institute of Chemistry, Chinese Academy of Science, Beijing~100190, P.R.~China
}

\email{\authormark{*}florian.vogelbacher.fl@ait.ac.at} %% email address is required

% \homepage{http:...} %% author's URL, if desired

% Style guide notes
% Equations: use Eq. (1), Eq. (2), etc.

%%%%%%%%%%%%%%%%%%% abstract %%%%%%%%%%%%%%%%
%% [use \begin{abstract*}...\end{abstract*} if exempt from copyright]

\begin{abstract}
The optimization of material gain in optically pumped dye-doped polymer thin films is an important task in the development of organic solid-state lasers. In this work, we present a theoretical model that accommodates the influence of concentration quenching on material gain and employ it to study the novel dye molecule 2-(4-(bis(4-(\textit{tert}-butyl)\-phenyl)\-amino)\-benzylidene)\-malononitrile (PMN) and the well-established dye molecule 4-(di\-cyano\-methylene)-2-methyl-6-(4-dimethylaminostyryl)-4H-pyran (DCM) embedded in poly(methyl~methacrylate) (PMMA). Polycarbonate was tested as an alternative host material for PMN. The material gain in these dye-doped polymer thin films was determined by the variable stripe length method. The inclusion of concentration quenching in the material gain expression is able to significantly reduce the overestimation of the gain efficiency inherent to a linear model.
\end{abstract}

%%%%%%%%%%%%%%%%%%%%%%%%%%  Introduction  %%%%%%%%%%%%%%%%%%%%%%%%%%
\section{Introduction}
Photonic technology based on polymer materials provides a versatile platform for applications in data- \& telecommunication, sensing, and medical diagnostics, achievable by a broad range of material properties, flexibility in fabrication, and low cost \cite{Ma.2002b, Wang.2012i, Kim.2013, Ostroverkhova.2013}. Suitable optical and optoelectronic characteristics enabled the development of organic solar cells \cite{Mayer.2007, Gunes.2007} and highly efficient light sources, such as organic solid-state lasers \cite{Forget.2013} and light emitting diodes~(OLEDs) \cite{Tang.1989, Nakanotani.2014, Sekine.2014}. In contrast to OLEDs, the feasibility of direct electrical pumping of organic solid-state lasers and continuous wave operation are still an open research topic. Therefore, organic solid-state lasers are driven by pulsed optical excitation \cite{Costela.2003, Samuel.2004, Forget.2013}. Different optical pump sources such as inorganic solid-state lasers, semiconductor laser diodes \cite{Yang.2008b, Foucher.2014, Zhao.2015}, or light-emitting diodes (LEDs) \cite{Yang.2008b, Herrnsdorf.2013} have been successfully employed \cite{Kuehne.2016}. Due  to their compact size and comparatively low cost semiconductor laser diodes and LEDs are particularly promising pump sources for applications in organic solid-state lasers in hand-held and point-of-care devices.  Even though the lifetime of organic dyes under strong optical excitation is limited due to photobleaching \cite{Beer.1972, McKenna.2004, Mondal.2008}, their use in low cost sensing applications and single-use devices might be favorable compared to more expensive semiconductor devices. The possibility to locally deposit the dye-doped polymer material on the substrate via ink-jet printing can decrease fabrication costs even further and adds additional design flexibility \cite{Bollgruen.2016, Alaman.2016}. Compatibility to CMOS fabrication processes enables the co-integration of organic polymer based technology with microelectronics. 

Small molecular dyes embedded in transparent host materials are, in general, a popular class of organic gain materials for organic dye lasers due to their wide spectral emission properties ranging from the near ultra violet to near infrared wavelength region. Early works investigated the influence of dye concentration, excitation intensity, and pulse width on lasing properties of liquid dye lasers\cite{Wieder.1972, Speiser.1973}. However, for thin film polymer solid-state lasers concentration dependent lasing properties have not been widely reported despite the strong influence of the doping level and the host matrix \cite{Green.2015, AmyotBourgeois.2017, MunozMarmol.2018}. In particular, the evaluation of the material gain is a prerequisite for the development and proper choice of dye molecules as well as the selection of suitable host polymers, as concentration quenching ultimately limits attainable modal gain. 

In this work, we present a model for material gain concentration quenching. We tested it against gain properties of the novel electron donor-electron acceptor dye molecule 2-(4-(bis(4-(\textit{tert}-butyl)phenyl)amino)benzylidene)malononitrile (PMN), which was designed for improved intramolecular charge transfer and broad spectral absorption in the visible region, and the commercially available dye molecule 4-(dicyanomethylene)-2-methyl-6-(4-dimethylaminostyryl)-4H-pyran (DCM), which represents a good reference due to its widespread use as a gain material. The influence of the host polymer matrix on the quenching was investigated by embedding PMN in poly(methyl~methacrylate)~(PMMA) and polycarbonate~(PC). The results were compared against the proposed quenching model. Gain properties of dye-doped PC thin films have not yet been published widely, despite its broad use in light guiding, optics, optical data storage and light sources as an highly transparent plastic \cite{Baumer.2010}. The higher refractive index of PC compared to PMMA is an interesting material property for integrated photonic designs. The modal gain of the dye-doped polymer thin films was determined by the variable stripe length (VSL) method \cite{Shaklee.1971, Negro.2004, Cerdan.2010, Gozhyk.2015}. Optical excitation of the gain material was performed by a high power \SI{450}{\nano\meter} blue laser diode enabling a material characterization at the exact wavelength, optical intensity, and temporal pulse shape of the envisaged light source.

%%%%%%%%%%%%%%%%%%%%%%%%%%  Materials and Method  %%%%%%%%%%%%%%%%%%%%%%%%%%
\section{Materials and method}
%-------------------------
\subsection{Organic dyes}
%-------------------------
For the synthesis of the novel dye, freshly distilled \ch{POCl3} (\SI{28.4}{\milli\liter}, $12.5$~eq) was added dropwise to dimethylformamid (DMF) (\SI{21.7}{\milli\liter}, $12$~eq) under an atmosphere of \ch{N2} at \SI{0}{\celsius} in ice bath, and then it was stirred for \SI{1}{\hour}. \textit{N},\textit{N}-Di(4-\textit{tert}-butylphenyl)aniline (\SI{4.3}{\gram}, \SI{12}{\milli\mol}) was added to the above solution, and the resulting mixture was stirred for \SI{4}{\hour} at \SI{0}{\celsius}. Then, the mixture was poured into a beaker containing ice-cubes, and basified with \SI{4}{\Molar} \ch{NaOH}. The solid was filtered and extracted with ethylacetate (EA)-brine. After evaporating the organic solvent, the crude product was purified by column chromatography on silica using a mixture of EA-hexane (1:6, v/v), to give \SI{4.1}{\gram} of a yellow solid \cite{Klimavicz.2012}. Under an atmosphere of dry nitrogen, 4-[Bis[4-(1,1-dimethylethyl)phenyl]amino]benzaldehyde (\SI{2}{\gram}, \SI{5}{\milli\mol}) and malononitrile (\SI{0.6608}{\gram}, \SI{10}{\milli\mol}), dissolved in toluene (\SI{50}{\milli\liter}), were refluxed \SI{24}{\hour} using trace piperidine as the catalyst. Then, the reaction mixture was cooled to room temperature and the solvent was removed under reduced pressure. The resultant residue was purified by silica gel column chromatography using ethyl acetate and hexane (1:8, v/v) as eluents to afford the product as an orange-red solid (\SI{1.8}{\gram}, $83$\%), 2-(4-(bis(4-(\textit{tert}-butyl)phenyl)amino)benzylidene)malononitrile~(PMN). 

\begin{figure}[htbp]
\centering
\includegraphics{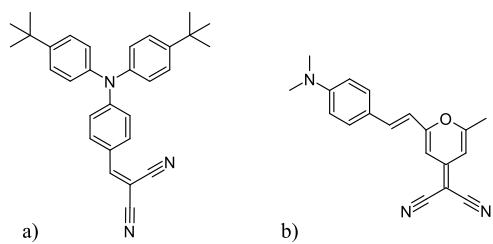}
\caption{Molecular structure of (a) 2-(4-(bis(4-(\textit{tert}-butyl)\-phenyl)\-amino)\-benzylidene)\-malononitrile~(PMN), and (b) 4-(dicyanomethylene)-2-methyl-6-(4-dimethylaminostyryl)-4H-pyran  (DCM).}
\label{fig:chemicalStructure}
\end{figure}

Figure~\ref{fig:chemicalStructure}(a) shows the molecular structure of the resulting dye. The tert-butyl decorated triphenylamine functions as donor and the cyanogroups act as acceptor. The bulky tert-butyl groups also improve the solubility, impede dye aggregation, and reduce concentration quenching. PMN shares a similar design to \emph{fvin}, an organic dye presented in \cite{Ishow.2008, RabbaniHaghighi.2009}. The commercially available benchmarking dye DCM, depicted in Fig.~\ref{fig:chemicalStructure}(b), was obtained from Sigma-Aldrich and was used without additional processing. Absorbance and fluorescence emission spectra of dye-doped polymer thin films on microscope cover slides were obtained by fluorescence spectroscopy (PerkinElmer, EnSpire). The excitation wavelength for the emission measurements was set to \SI{450}{\nano\meter} to be compatible with the operation wavelength of the pump laser diode used in the gain measurements later on.

%-------------------------
\subsection{Thin film preparation}
%-------------------------
Optical quality PMMA (Polycasa,  Acryl~G77, refractive index $n_\text{PMMA}=1.492$) granulate was dissolved in anisole (Sigma-Aldrich, $\geq99$\%) under magnetic stirring at \SI{30}{\celsius}, whereas 1,4-dioxane (Roth, $\geq99.5$\%) was used to prepare the PC (Covestro~AG, Makrolon LED2245, refractive index $n_\text{PC}=1.584$) solution. The corresponding dye was added to the solution after the polymer was completely dissolved. Different dye concentrations ranging from $0.5$~wt.\% to $5.0$~wt.\% have been obtained by mixing stock solutions. Since the solubility of DCM in 1,4-dioxane was limited, the material system DCM in PC was not deployed. The mass fractions of the dissolved polymers in solution were set to $10$~wt.\% for PC and $15$~wt.\% for PMMA, respectively, to achieve suitable viscosities for spin coating. Microscope cover slides (Menzel, \SI{25x25}{\milli\meter}) and silicon samples (\SI{20x20}{\milli\meter}) with a \SI{2}{\micro\meter} silicon dioxide ($n_\text{SiO2}=1.46$) layer were cleaned with an alkaline detergent solution (Hellma, Hellmanex) in an ultrasonic bath, thoroughly rinsed with deionized water, ultrasonically cleaned in acetone and isopropanol, and blow dried using nitrogen. Prior to spin coating, all substrates were dehydrated at \SI{200}{\celsius} for \SI{5}{\minute} on a hot plate. The higher refractive indices of the polymers compared to the silicon dioxide film on the silicon sample enable waveguiding for VSL measurements. PMMA thin films were prepared by spin coating in an open bowl configuration, resulting in thin film thickness of \SI{800 \pm 50}{\nano\meter}. For improved thin film quality of the PC resist, spin coating was conducted in a small volume closed bowl configuration flushed with nitrogen. A two step process was used to improve adhesion of the PC film to the substrate: In a first run, a \SI{15}{\nano\meter} thin non-doped PC layer was spin coated on the substrate and placed onto a hot-plate at \SI{200}{\celsius} for \SI{1}{\minute}, heating the samples above the polymer glass temperature. In a second step, the dye doped PC thin film was deposited onto the substrate, resulting in a total thin film thickness of \SI{650 \pm 50}{\nano\meter}. A baking step, common for both polymer types, at \SI{110}{\celsius} for \SI{1}{\minute} on a hot plate removed residual solvent without thermal bleaching the dye. Thin film thickness was determined through evaluation of the step height of locally removed polymer by stylus profilometry (KLA-Tencor, Alpha-Step IQ). The silicon samples for gain measurements were cleaved to obtain a sharp edge and to enable a measurement at a well defined thin film thickness at the center of the sample. 

%-------------------------
\subsection{Optical setup}
\begin{figure}[tbp]
\centering
\includegraphics{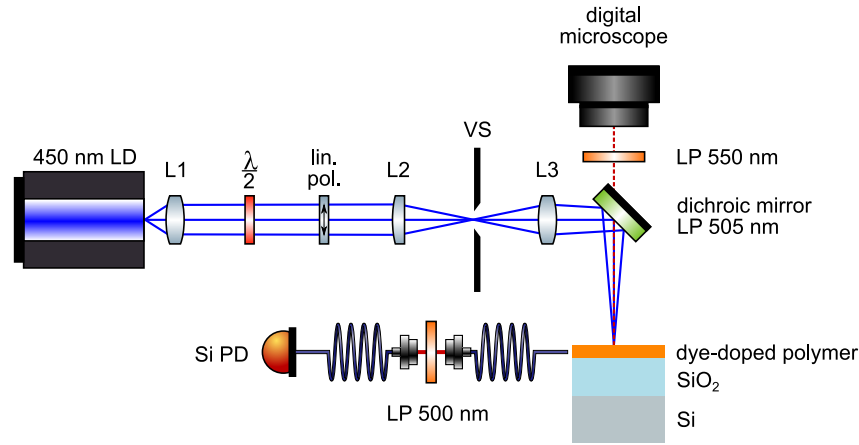}
\caption{Optical setup to generate a high intensity pump stripe on the sample surface. The intermediate image is used to add a variable slit (VS) minimizing Fresnel diffraction for variable stripe length (VSL) gain measurements. In this VSL setup no cylindrical lens is needed because the emission characteristics of the high power \SI{450}{\nano\meter} multimode laser diode directly results in a narrow pump stripe. The light emitted from the sample edge is coupled into a multimode fiber and detected by a silicon photodiode (Si PD). A \SI{500}{\nano\meter} long pass (LP) filter suppresses spurious pump light coupled into the fiber. Alignment of the pump stripe to the sample edge and fiber is achieved by inspection through a \SI{550}{\nano\meter} LP filter.}
\label{fig:pumpSetup}
\end{figure}

%-------------------------
\label{sec:optical_setup}
Short current pulses at a repetition rate of \SI{3}{\hertz} were generated by a pulse driver (PicoLAS, LDP-V~50-100) to drive a TO-can high power multimode \SI{450}{\nano\meter} laser diode (Nichia NDB7K75), reaching an electric current of up to \SI{40}{\ampere}. Figure~\ref{fig:pumpSetup} shows the optical setup to create a high intensity pump spot on the sample for VSL measurements. An aspheric lens~L1 collimated the strongly diverging laser diode beam. The combination of a half-wave plate and a linear polarizer allowed rotating the input beam polarization and attenuating the pump intensity without changing the laser diode drive current. A change in laser diode drive current would influence the optical pump pulse shape and thereby the gain measurements \cite{Herrnsdorf.2013}. Truncation of the pump beam for VSL measurements leads to Fresnel diffraction effects on the sample surface, which potentially degrade the quality of the gain measurements \cite{Negro.2004}. In our setup, we minimized these diffraction effects of the pump beam by placing a variable slit (VS) at an intermediate image created by lens~L2. The slit width was adjusted by a manual micrometer drive. In contrast to other optical setups, where a cylindrical lens was used to form a thin stripe \cite{Lu.2004b, Gozhyk.2015}, the large aspect ratio of the pump spot originated directly from the laser diode emission characteristics. The slit was imaged, deflected by a dichroic mirror, onto the sample surface by lens L3 creating a \SI{400}{\micro\meter} by \SI{20}{\micro\meter} stripe. Intensities of approximately \SI{150}{\kilo\watt\per\square\centi\meter} for \SI{35}{\nano\second} pulses were reached. A digital microscope was used to align the optical setup with the sample. Back reflected pump light from the sample surface, which spuriously passed the dichroic mirror, was blocked by a \SI{550}{\nano\meter} long pass filter. A multimode fiber ($0.10$~NA, \SI{10}{\micro\meter} core diameter) attached to a three axis flexure stage guided the light collected at the sample edge to a silicon photodiode (Si~PD) connected to a lock-in amplifier. A \SI{500}{\nano\meter} long pass filter mounted in a free space fiber-to-fiber bench suppressed pump light scattered and coupled into the fiber. The polarization of the pump light was set perpendicular to the long axis of the pump spot, resulting in a dominantly transverse electric (TE) amplified spontaneous emission into the stripe \cite{Gozhyk.2012}. 

\begin{figure}[tbp]
\centering
\includegraphics{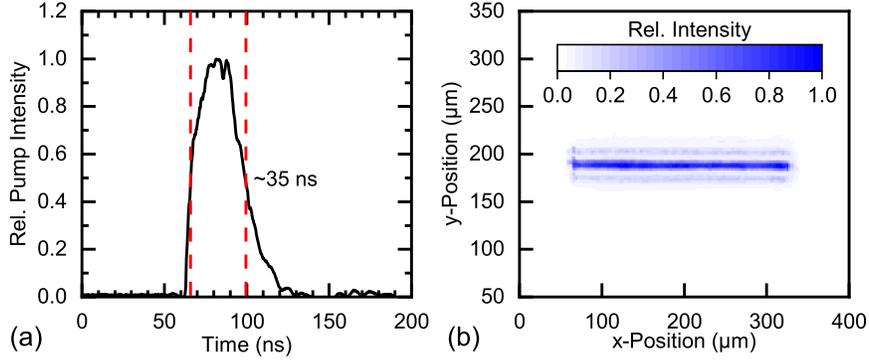}
\caption{(a) Temporal pump power characteristics of the laser diode driven by a pulsed current source. (b) Spatial intensity profile of the pump stripe. The optical setup suppresses Fresnel diffraction resulting in a sharp truncation of the pump stripe. Weak side lobes occur due to the emission characteristics of the LD pump source.}
\label{fig:pump_properties}
\end{figure}

The resulting temporal and spatial shapes of the optical pump setup are shown in Figs.~\ref{fig:pump_properties}(a) and \ref{fig:pump_properties}(b), respectively. The temporal behaviour was measured by a \SI{1}{\giga\hertz} silicon photodiode connected to an oscilloscope with \SI{1}{\giga\hertz} bandwidth. The full width at half maximum (FWHM) of the optical pulse was \SI{35}{\nano\second} at a peak driving current of \SI{38}{\ampere}. The spatial pump intensity variation was measured to be less than $20$\% along the pump stripe and diffraction effects of the variable slit were efficiently suppressed. Side lobes of \SI{-6}{\decibel} compared to the main pump stripe intensity originating from the laser diode emission characteristics occurred symmetrically to the main pump stripe. These weak side lobes are not expected to detrimentally influence the gain measurements.

%---------------------------------------------------------------------
\subsection{Variable stripe length method}
%---------------------------------------------------------------------
\label{sec:variable_stripe_length_method}
Amplified spontaneous emission (ASE) measurements in a variable stripe length configuration are commonly employed to determine the modal gain $g_\text{mod}$ of polymer thin film gain media, which can subsequently be used to extract the material gain. To account for geometrical effects of the narrow pump stripe and the influence of the light collecting fiber, but before intensity dependent saturation effects arise, an expression for the amplification of spontaneous emission is presented in the following.

\begin{figure}[htb]
\centering
\includegraphics{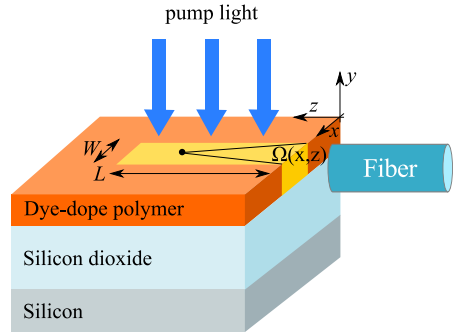}
\caption{Schematic drawing of the fiber-coupled variable stripe length (VSL) method to measure modal gain. The pump stripe has a length $L$ and width $W$. The solid angle of the fluorescence emission at position $(x,z)$ subtended by the pump stripe output facet is denoted with $\Omega(x,z)$.}
\label{fig:variable_stripe_length}
\end{figure}

Optical excitation of the embedded dye molecules leads to a local fluorescence emission density $I'_\text{fl}(x,z)$. In absence of amplification, the contribution to the light collected by the fiber at a given $z$-position can be written as the integral across the pump stripe resulting in an effective fluorescence density 
\begin{equation}
\label{eqn:fluorescence_integral}
I'_\text{fl,eff}(z) = \int^{W/2}_{-W/2} I'_\text{fl}(x,z) \frac{\Omega(x,z)}{2\pi} \gamma(x,z)\;\textrm{d}x,
\end{equation}
with the pump stripe width $W$, the solid angle $\Omega(x,z)$ and the fiber coupling efficiency $\gamma(x,z)$. The solid angle depicted in Fig.~\ref{fig:variable_stripe_length} is defined by the position of the emitting fluorophore and the pump stripe exit aperture. The quantity ${\Omega(x,z)}/{2\pi}$ represents the fraction of spontaneous emission propagating along the pump stripe to the exit aperture. For increasing separation between the local fluorescence emission and the sample edge, the effective fluorescence density converges to $I'_\text{fl,eff}(z) \propto z^{-1}$. The fact that the coupling efficiency to the fiber depends on the position of the fluorophore to the end facet is included in this model through the coupling efficiency factor $\gamma(x,z)$. This correction factor is influenced by the numerical aperture and core diameter of the fiber, and the pump stripe width. To a good approximation a geometry dependent parameter $z_0$ can be introduced, which accounts for all these geometrical influences. In the following, the effective fluorescence density is assumed to follow the relation
\begin{equation}
\label{eqn:coupling_model}
I'_\text{fl,eff}(z) \propto \frac{1}{z+z_0}.
\end{equation}
The validity of Eq.~\eqref{eqn:coupling_model} as a suitable model for a fiber-coupled pump stripe geometry was verified through numerical calculations and confirmed by measurement data. 

In the presence of gain the change of light intensity $I_\text{ASE}$ along the pump stripe in a slab waveguide thin film system can be written as differential equation (cf.\@ \cite{Negro.2004, Milonni.2010}) of the form
\begin{equation}
\label{eqn:diff_equation_ASE}
\diff{I_\text{ASE}}{z} = g_\text{mod}I_\text{ASE} +  I'_\text{fl,eff}(z).
\end{equation}
The solution of the differential equation with the emission coupling model in Eq.~\eqref{eqn:coupling_model} and satisfying the boundary condition $I_\text{ASE}(0)=0$, results in
\begin{equation}
\label{eqn:diff_equation_ASE_solution}
I_\text{ASE}(L) = A \exp\left( g_\text{mod}(L+z_0) \right) \left[ \expintegral(-g_\text{mod}(L+z_0)) - \expintegral(-g_\text{mod}z_0) \right],
\end{equation}
where $\expintegral$ denotes the \emph{exponential integral} defined as 
\begin{equation}
\label{eqn:exponential_integral}
\expintegral(z) = \int_{-z}^{\infty} \frac{\exp(-t)}{t}\;\textrm{d}t.
\end{equation}
The coefficient $A$ in Eq.~\eqref{eqn:diff_equation_ASE_solution} is a fluorescence intensity coefficient. Without taking the correction term ${\Omega(x,z)}\gamma(x,z)/{2\pi}$ into account, the amount of fluorescence light contributing to $I_\text{ASE}$ would be overestimated. As a consequence, $g_\text{mod}$ derived from the measured VSL data would be underestimated. This becomes particularly apparent for low dye concentrations, where high negative gain would be derived from the data.

In the limit of a long pump stripe the solid angle $\Omega(x,z)$ can be assumed constant. The fraction of spontaneous emission contributing to ASE results in $\Omega/2\pi \approx W/L$. If the coupling to the light collecting optics is independent of the pump stripe output emission angle, the coupling efficiency $\gamma$ is constant. In summary, this results in the exponential dependence of the ASE intensity \cite{Negro.2004}
\begin{equation}
\label{eqn:ASE}
I_\text{ASE}(L) = \frac{A}{g_\text{mod}} \left( e^{g_\text{mod}L}-1 \right),
\end{equation}
as a solution of Eq.~\eqref{eqn:diff_equation_ASE}. Equation~\eqref{eqn:ASE} is commonly used for optical pump setups where long pump stripes are realized and geometric effects do not influence the gain measurements \cite{Negro.2004,Cerdan.2010}. However, in this work the laser diode driven VSL optical setup required the use of the more general result from Eq.~\eqref{eqn:diff_equation_ASE_solution} to obtain valid gain values. For reliable gain measurements the pump stripe length should not exceed the saturation length $L_\text{sat}$ of the studied system \cite{Negro.2004, Cerdan.2010}, which is fulfilled for the maximum pump length and gain materials in this work. Exemplary VSL curves are presented in section \ref{sect:results_and_discussion}, Fig.~\ref{fig:gain_efficiency}(a). For ASE intensities reaching the saturation level, Eq.~\eqref{eqn:diff_equation_ASE_solution} is no longer valid and has to be refined to account for saturation effects \cite{Costela.2008, Costela.2008b}.

The modal gain $g_\text{mod}$ can be broken down into additional terms to distinguish between different contributions, such that \cite{Negro.2004}
\begin{equation}
\label{eqn:modal_gain}
g_\text{mod} = \Gamma g_\text{mat} - \alpha_\text{mod},
\end{equation}
where $\Gamma$ denotes the confinement factor of the thin film mode, and $\alpha_\text{mod}$ is a loss term that includes modal losses of an undoped thin film system, e.g.\@ surface scattering or power leakage into the substrate. In contrast, polymer material loss and dye doping dependent losses are included in the material gain expression through a material loss term $\alpha_\text{mat}$, which yields
\begin{equation}
\label{eqn:material_gain}
g_\text{mat} = g_\text{int} - \alpha_\text{mat}.
\end{equation}
The material loss $\alpha_\text{mat}$ includes re-absorption and scattering of the fluorescence light, and $g_\text{int}$ is a bulk material internal gain. Equation~\eqref{eqn:modal_gain} can be rewritten as
\begin{equation}
\label{eqn:modal_gain_2}
g_\text{mod} = \Gamma g_\text{int} - \alpha,
\end{equation}
where the loss $\alpha$ represents both material and modal losses 
\begin{equation}
\label{eqn:alpha}
\alpha = \Gamma \alpha_\text{mat} + \alpha_\text{mod}
\end{equation}
and is dependent on the doping level of dye molecules in the host polymer.

%-----------------------------------------
\subsection{Concentration quenching model}
%-----------------------------------------
\label{sect:concentration_quenching}
Concentration quenching in a gain material is a parasitic effect that occurs when the dye concentration in the host material is increased beyond a critical value above which the average distance between dye molecules becomes so small that intermolecular interactions start to adversely affect the dye properties. Concentration quenching reduces the fluorescence quantum yield $\eta_\text{QY}$ of the gain material and, consequently, limits the performance. In the following, we elaborate a model that allows expressing the material gain $g_\text{mat}$ as a function of the dye concentration taking into account concentration quenching. 

An exponential dependence of $\eta_\text{QY}$ on the average molecular separation $r$ between dye molecules embedded in PMMA has been observed by Green et al. \cite{Green.2015}. Therefore, we model the fluorescence quantum yield, using the relation $r \propto N^{-1/3}$, as
\begin{equation}
\label{eqn:quantum_yield}
\eta_\text{QY}(N) = \eta_0\left(1-e^{-{\beta'}^{-1/3} \left(N^{-1/3}-N_0^{-1/3}\right)}\right),
\end{equation}
with the molecular densities of the dye $N$ and $N_0$, the quenching strength $\beta'$, and the low concentration, i.e.\@ non-quenched, quantum yield constant $\eta_0$. The parameter $N_0$ represents the molecular density at which the quantum yield vanishes. By definition, the quantum yield is the ratio between fluorescence lifetime $\tau_\text{f}$ and radiative lifetime $\tau_\text{rad}$, namely
\begin{equation}
\label{eqn:quantum_yield_def}
\eta_\text{QY}(N)=\frac{\tau_\text{f}(N)}{\tau_\text{rad}},
\end{equation}
where the radiative decay time $\tau_\text{rad}$ is assumed to be independent of the molecular dye density. On the other hand, the fluorescence lifetime $\tau_f(N)$ is defined as the inverse of the sum of the radiative, non-radiative, and concentration quenching decay rates $k_\text{R}$, $k_\text{NR}$, $k_\text{CQ}$\cite{Green.2015}
\begin{equation}
\label{eqn:fluorescence_lifetime}
\tau_f(N)=\frac{1}{k_\text{R}+k_\text{NR}+k_\text{CQ}(N)}.
\end{equation}
Therefore, the quantum yield becomes dependent on the molecular density of the embedded dye in the host matrix through the concentration quenching decay rate. To estimate the influence of $\eta_\text{QY}(N)$ on the population inversion, the steady state rate equation, neglecting intersystem crossing (ISC) or additional higher order decay paths, 
\begin{subequations}
\begin{align}
\label{eqn:rate_equation}
\frac{dS_1}{dt}  = \sigma_\text{abs}I_\text{p} S_0 - \left(\frac{1}{\tau_f}+\sigma_\text{em}I\right)S_1 & = 0 \\
	S_0 + S_1	 & = 1
\end{align}
\end{subequations}
with the relative singlet state populations $S_1$ and $S_0$, the pump photon density $I_\text{p}$, the emission photon density $I$, and the absorption cross section $\sigma_\text{abs}$, is solved for $S_1$ resulting in
\begin{equation}
\label{eqn:population_inversion}
S_1 = \frac{\tau_f \sigma_\text{abs} I_\text{p}}{1+\tau_f(\sigma_\text{em}I+\sigma_\text{abs}I_\text{p})} \approx \tau_f \sigma_\text{abs} I_\text{p}.
\end{equation}
The approximation used in Eq.~\eqref{eqn:population_inversion} is only valid in a small signal approximation $I \ll I_p$ and for low to moderate pump intensities where saturation effects can still be neglected such that the singlet state population $S_1$ depends linearly on the pump photon density. For typical values of organic dyes, which amount to $\sigma_\text{abs}\approx\SI{1e-16}{\square\centi\meter}$ and $\tau_f\approx \SI{2}{\nano\second}$ \cite{Forget.2013}, and a pump intensity of \SI{150}{\kilo\watt\per\square\centi\meter}, the deviation is below $10\%$. At high pump intensities, saturation effects limit the relative singlet state population to $S_1<1$ and the approximation in Eq.~\eqref{eqn:population_inversion} is no longer valid.

The internal material gain $g_\text{int}$ used in Eq.~\eqref{eqn:material_gain} to express the material gain $g_\text{mat}$ depends linearly on the emission cross section $\sigma_\text{em}$ and the population inversion density $\Delta N$ between the upper and lower level of the lasing transition, such that 
\begin{equation}
\label{eqn:internal_material_gain_01}
g_\text{int} = \sigma_\text{em}\Delta N. 
\end{equation}
For organic dyes the non-radiative relaxation from the lower fluorescence level to the ground state $S_0$ is usually a very fast decay path. This allows approximating the population inversion by $\Delta N\approx S_1 N$. Employing  Eq.~\eqref{eqn:quantum_yield_def} and Eq.~\eqref{eqn:population_inversion} the internal material gain can then be expressed as a function of the molecular density of the dye $N$, i.e.\@ 
\begin{equation}
\label{eqn:internal_material_gain_02}
	g_\text{int}(N) \approx \sigma_\text{em} S_1 N = \eta_\text{QY}(N) \tau_\text{rad} \sigma_\text{em} \sigma_\text{abs} I_\text{p} N .
\end{equation}
The internal material gain, using the quantum yield expression from Eq.~\eqref{eqn:quantum_yield}, can be written as
\begin{equation}
\label{eqn:internal_material_gain_quenched_molecular}
g_\text{int}(N) = \eta_0\left(1-e^{-\beta^{-1/3}\left(N^{-1/3}-N_0^{-1/3}\right)}\right) \tau_\text{rad} \sigma_\text{em} \sigma_\text{abs} I_\text{p} N.
\end{equation}
Expressing the dye content in the thin film through the weight ratio 
\begin{equation}
w \approx \frac{N \cdot M}{\rho \cdot N_\text{A}},
\end{equation}
with the polymer density $\rho$, the molar mass $M$ of the dye, and the Avogadro constant $N_\text{A}$, Eq.~\eqref{eqn:internal_material_gain_quenched_molecular} can be formulated as a function of the weight ratio of the dye 
\begin{equation}
\label{eqn:internal_material_gain_quenched_fractional}
g_\text{int}(w) = K(w) I_\text{p}
\end{equation}
with the mass fraction dependent gain efficiency
\begin{equation}
\label{eq:gain_efficiency_quenched}
K(w) = w \frac{\rho \cdot N_\text{A}}{M}\eta_0\left(1-e^{-\beta^{-1/3}\left(w^{-1/3}-w_0^{-1/3}\right)}\right) \tau_\text{rad} \sigma_\text{em} \sigma_\text{abs}.
\end{equation}
The quenching strength $\beta$ in Eq.~\eqref{eq:gain_efficiency_quenched} differs from $\beta'$ in Eq.~\eqref{eqn:quantum_yield} by a constant factor arising out of the change from the molecular density $N$ to the mass fraction $w$. Based on Eq.~\eqref{eqn:internal_material_gain_quenched_fractional} and Eq.~\eqref{eqn:modal_gain_2} the modal gain $g_\text{mod}$ results in
\begin{subequations}
\begin{align}
g_\text{mod}(w) & = \Gamma g_\text{int}-\alpha \\
    & = \Gamma K(w) I_\text{p} - \alpha.
\end{align}
\end{subequations}
Since we are investigating dye doped polymer thin films at low to moderate doping levels, the vertical reduction of the pump light from absorption is neglected and the pump photon density $I_\text{p}$ is assumed constant inside the gain material. A spatial variation of the pump intensity leads to a local variation of gain. In particular, pump diffraction effects can cause artificial gain \cite{Negro.2004}. Special care was taken in the design of the optical setup to minimize these effects by placing the variable slit at the position of an intermediate image.

%%%%%%%%%%%%%%%%%%%%%%%%%%  Results and Discussion  %%%%%%%%%%%%%%%%%%%%%%%%%%
\section{Results and discussion}
\label{sect:results_and_discussion}
%-----------------------------------------
\subsection{Absorbance and emission}
\begin{figure}[htbp]
\centering
\includegraphics{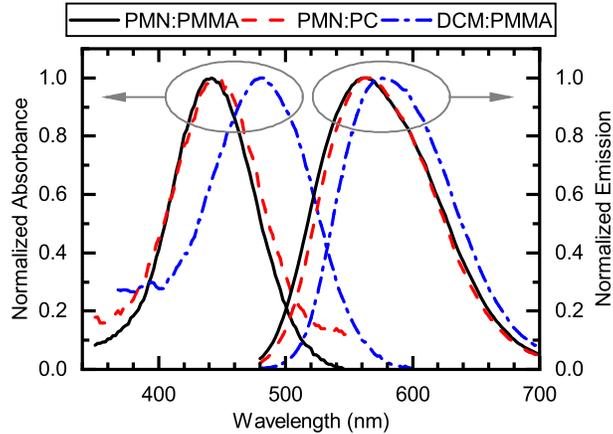}
\caption{Normalized absorbance and emission spectra for PMN ($2.0$~wt.\%) in PC, and PMN ($2.0$~wt.\%) and DCM ($1.0$~wt.\%) in PMMA. Excitation wavelength for fluorescence emission spectra was set to \SI{450}{\nano\meter}, corresponding to the operation wavelength of the pump laser diode used in gain measurements.}
\label{fig:absorbance_emission}
\end{figure}

%-----------------------------------------
Absorbance and emission spectra of $2.0$~wt.\% PMN in PMMA and PC, and $1.0$~wt.\% DCM embedded in PMMA are shown in Fig.~\ref{fig:absorbance_emission}(a) and Fig.~\ref{fig:absorbance_emission}(b), respectively. The absorbance peaks of PMN and DCM are located at \SI{440}{\nano\meter} and \SI{480}{\nano\meter}, respectively. Efficient absorption at the laser diode pump wavelength is expected for both dyes, as the excitation wavelength of \SI{450}{\nano\meter} is close to the absorbance peaks. Peak emission wavelengths for PMN and DCM are \SI{564}{\nano\meter} and \SI{578}{\nano\meter}, respectively. The large Stokes shifts of \SI{124}{\nano\meter} for PMN and \SI{98}{\nano\meter} for DCM reduce the influence of re-absorption. Only a minor red shift was observed for PMN in PC films with respect to PMN in PMMA films.

%-----------------------------------------
\subsection{Material gain and concentration quenching}
\begin{figure}[htbp]
\centering
\includegraphics{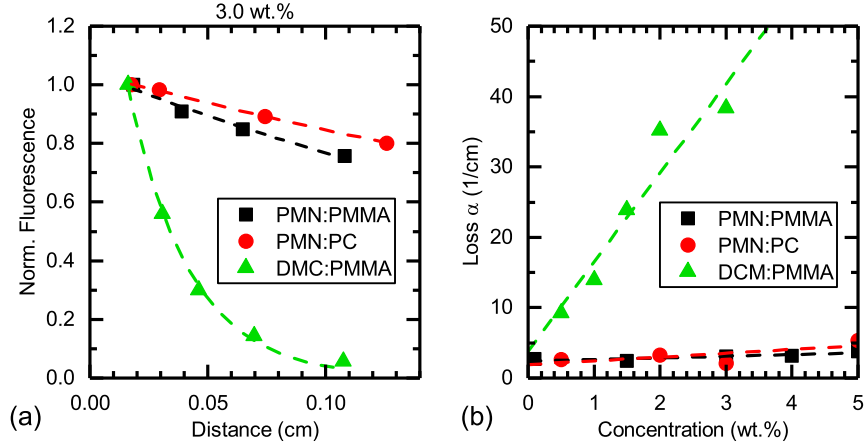}
\caption{(a) Dependence of fluorescence intensity on distance of excitation stripe from the sample edge. The pump stripe was orientated parallel to the sample edge to minimize the influence of the fiber NA on the measurement. Dashed lines present exponential fit to the data. (b) Derived losses $\alpha=\Gamma \alpha_\text{mat}+\alpha_\text{mod}$ from shifting excitation stripe measurements. Dashed lines present a linear fit to the loss.}
\label{fig:loss_total}
\end{figure}

%-----------------------------------------
To obtain valid material gain data for a concentration quenching analysis, the geometrical correction described in section \ref{sec:variable_stripe_length_method} had to be taken into account to accommodate the effect of the pump stripe geometry and fiber coupling efficiency present in our VSL measurement setup. In a first step, the concentration dependent loss $\alpha$ was determined by shifting an excitation stripe orientated parallel to the sample edge. A parallel orientation of the excitation stripe effectively reduced the influence of mode spreading and of the NA of the collecting fiber. An exponential fit to the obtained intensity data was used to extract the loss $\alpha$, shown in Fig.~\ref{fig:loss_total}(a), which was repeated for various dye concentrations of PMN and DCM in the two host polymer materials. Subsequently, the modal loss $\alpha_\text{mod}$ was derived from the intercept with the y-axis of a linear data fit to the loss $\alpha$ and amounted to less than \SI{3}{\per\centi\meter} for both PMMA and PC. Loss increased moderately for increasing dye concentrations in PMN doped polymers, whereas a strong rise in loss was measured for DCM. The significantly higher loss of DCM compared to PMN could be a result of absorbing impurities present in the employed DCM dye and re-absorption of the fluorescence light in the emission wavelength region despite the strong Stoke's shift. In general, high loss reduces modal gain and no light amplification occurs at elevated dye concentrations. Results of the performed loss measurements are summarized in Fig.~\ref{fig:loss_total}(b), where the linear fits to the determined losses for different dye concentrations are presented.

The fitting parameter $z_0$, which is required to model the ASE intensity using Eq.~\eqref{eqn:diff_equation_ASE_solution}, was extracted from measurements at low dye concentrations ($0.1$~wt.\%), for which ASE becomes negligible, i.e.\@ vanishing internal gain, and spontaneous emission dominates. Subsequently, the parameter $z_0$ and the determined loss values were used to extract the internal gain from the VSL data at different dye concentrations of PMN and DCM in the corresponding polymer material systems. The internal gain is related to the modal gain through Eq.~\eqref{eqn:modal_gain_2}.

\begin{figure}[htbp]
\centering
\includegraphics{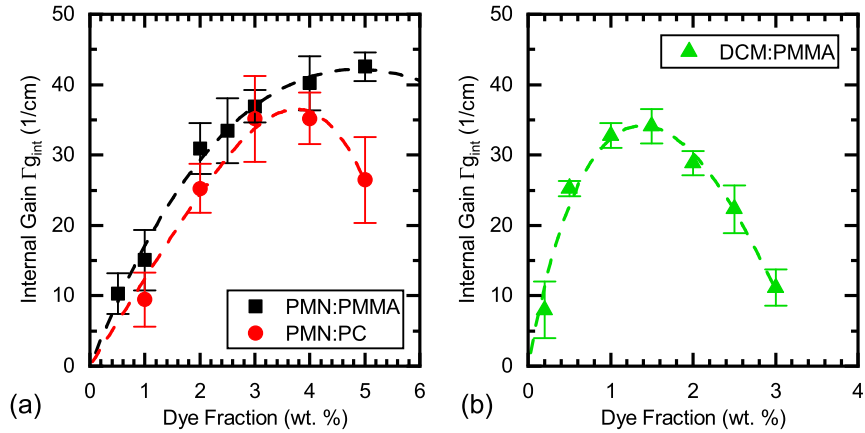}
\caption{Dependence of internal material gain $g_\text{int}$ on dye mass fraction for different dye-doped thin film polymers: (a) PMN:PMMA, PMN:PC, (b) DCM:PMMA. Measurements were carried out at \SI{150}{\kilo\watt\per\square\centi\meter} pump intensity and \SI{35}{\nano\second} pulse width. Graphs include best fits to the proposed concentration quenching gain model, shown as dashed lines.}
\label{fig:gain_curve}
\end{figure}

For PMN, a peak internal gain $\Gamma g_\text{int}$ of \SI{43 +- 2}{\per\centi\meter} was obtained at a concentration of $5.0$~wt.\% in PMMA, whereas a peak internal gain of \SI{35 +- 6}{\per\centi\meter} at a concentration of $4.0$~wt.\% was reached in PC. The influence of concentration quenching on the internal gain  was more pronounced in PC than in PMMA. For DCM in PMMA the peak internal gain of \SI{34 +- 3}{\per\centi\meter} occurred at a significantly lower concentration of $1.5$~wt.\%. Figures~\ref{fig:gain_curve}(a) and \ref{fig:gain_curve}(b) summarize the internal gain $\Gamma g_\text{int}$ characteristics of both PMN and DCM, including best fits to Eq.~\eqref{eqn:internal_material_gain_quenched_fractional} on the measured gain. The obtained quenching parameters are summarized in Table \ref{tab:quenching_parameters}.

\begin{table}[h!]
\caption{Concentration quenching parameters derived from gain measurements.}
\label{tab:quenching_parameters}
\centering
 \begin{tabular}{l l r r} 
 \hline
 Dye & Polymer & $w_0$ (wt.\%) & $\beta^{-1/3}$ ($\text{wt.\%}^{1/3}$) \tabularnewline
 \hline
 PMN & PMMA & $11 \pm 2$ & $4 \pm 2$ \tabularnewline
 PMN & PC & $6.0 \pm 0.5$ & $17 \pm 7$ \tabularnewline
 DCM & PMMA & $3.4 \pm 0.1 $ & $2 \pm 1$ \tabularnewline
 \hline
 \end{tabular}
\end{table}

The quenching concentration $w_0$, which describes the dye concentration for which the internal gain ultimately vanishes due to dye molecule intermolecular interaction, was highest for PMN in PMMA with $(11 \pm 2)$~wt.\%, suggesting a good performance at high concentrations. Concentration quenching set in at lower concentrations in the PC host polymer, with a model parameter $w_0$ of $(6.0 \pm 0.5)$~wt.\%. The quenching concentration of DCM molecule in PMMA was calculated to be $(3.4 \pm 0.1)$~wt.\%. The bulky tert-butyl groups of PMN therefore not only improve the solubility of the molecule but additionally reduce dye aggregation and concentration quenching. All measured systems are well described by the proposed concentration quenching model.

While the internal gain $\Gamma g_\text{int}$ describes the underlying quenching phenomena, modal gain includes inherent losses and is therefore an important figure to describe the net gain of the studied thin film system. A peak modal gain of \SI{39}{\per\centi\meter} was reached for PMN embedded in PMMA at a dye concentration of $5.0$~wt.\%, whereas \SI{32}{\per\centi\meter} was obtained for the PC host polymer at $3.0$~wt.\%. A peak modal gain of \SI{16}{\per\centi\meter} was determined for $1.0$~wt.\% DCM embedded in PMMA. Above $2.0$~wt.\%, no positive modal gain could be reached since quenching reduced the internal gain below the loss $\alpha$. 

In contrast to our findings, a modal gain of up to \SI{40}{\per\centi\meter} was reported for concentrations of DCM as high as $5.0$~wt.\% by Gozhyk et al.\@ \cite{Gozhyk.2015}. The difference can most likely be attributed to the high loss present in the tested DCM samples. Moreover, optical pumping was performed at a pulse width of \SI{500}{\pico\second} compared to \SI{35}{\nano\second} in our experiments. Such a short high intensity pulse width leads to an increased population inversion, which potentially reduces losses induced from higher order decay paths such as non-radiative intrinsic decay, intersystem crossing (ISC) to the triplet state, or molecule-molecule interaction based on F\"orster energy transfer. The temporal shape of the optical excitation has a significant influence on the determined gain and a direct comparison of gain values between different pump sources is in general difficult \cite{Herrnsdorf.2013}. Additionally, DCM is quite sensitive to intermolecular interaction, such that typical dye concentrations do not exceed $2.0$~wt.\%. To overcome this limitation, Vembris et al. added bulky trityloxyethyl groups to the molecule and successfully impeded quenching for these DCM derivatives \cite{Vembris.2017}.

\begin{figure}[htbp]
\centering
\includegraphics{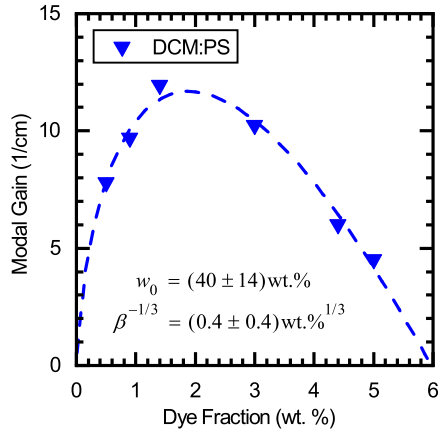}
\caption{Dependence of modal $g_\text{mod}$ on doping concentration of DCM embedded in polystyrene (PS). Modal gain data were taken from the work of Lu et al. \cite{Lu.2004b}. Excitation was performed at a wavelength of \SI{532}{\nano\meter} and a pulse width of \SI{8}{\nano\second}. The dashed line represents the best fit of the proposed concentration quenching gain model. Adapted with permission from \cite{Lu.2004b}, \emph{The Optical Society} (OSA).}
\label{fig:gain_curve_DCM_PS}
\end{figure}

To further verify the validity of our proposed concentration quenching gain model we applied it to gain data of a DCM doped polystyrene (PS) thin film system reported by Lu et al.\@ \cite{Lu.2004b}. The modal gain values of DCM:PS films have been obtained by Lu et al.\@ employing a Q-switched solid-state {Nd:YAG} laser working at a wavelength of \SI{532}{\nano\meter}. The pulse width was \SI{8}{\nano\second} at a repetition rate of \SI{10}{\hertz}. The loss amounted to \SI{9.25}{\per\centi\meter} for a concentration of $1.4$~wt.\%. On the basis of this values, we estimated the dye concentration dependent loss $\alpha$ using the linear model 
\begin{equation}
\label{eq:internal_gain_DCM_PS}
\alpha(w) = w \frac{\SI{9.25}{\per\centi\meter}}{1.4\;\text{wt.\%}},
\end{equation}
where the contribution of the modal loss $\alpha_\text{mod}$ was neglected.
Figure~\ref{fig:gain_curve_DCM_PS} shows the reported modal gain values of DCM in PS including a best fit to the quenching model and the derived quenching model parameters $\beta^{-1/3}$ and $w_0$. The high quenching concentration of $40$~wt.\% suggests a good performance of DCM in the PS host matrix. However, the high losses of DCM reduce the attainable modal gain. An extrapolation of the fit indicates that the net gain vanishes beyond a concentration of $6$~wt.\%. Overall, the modal gain of DCM in PS is well described by the concentration quenching model. 

%-----------------------------------------
\subsection{Gain efficiency of DCM}
%-----------------------------------------
Knowledge about the gain efficiency $\Gamma K$ facilitates the design of organic solid-state lasers, as the minimum pump intensity required for lasing can be estimated from the threshold gain of the laser cavity. For a better physical insight we determined the gain efficiency of DCM experimentally and compared it to both the commonly employed linear model and to the quenched model introduced in section \ref{sect:concentration_quenching}. For DCM the required material parameter values are well studied and available in literature.

\begin{figure}[htbp]
\centering
\includegraphics{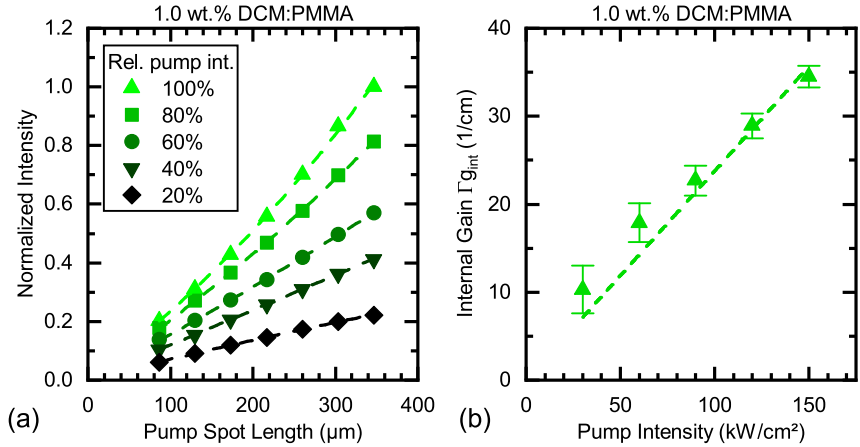}
\caption{(a) Normalized ASE intensity obtained from VSL gain measurements of $1.0$~wt.\% DCM in PMMA for different pump intensities relative to \SI{150}{\kilo\watt\per\square\centi\meter} at \SI{35}{\nano\second} pulse width. (b) Dependence of internal gain $\Gamma g_\text{int}$ on the pump intensity. The slope of a linear fit to the measured data yields the gain efficiency $\Gamma K$.}
\label{fig:gain_efficiency}
\end{figure}

VSL measurements were used to experimentally investigate the dependence of the internal gain $\Gamma g_\text{int}$ on the pump intensity. The gain efficiency $\Gamma K$ was extracted from the slope of a linear fit to the pump intensity dependent 
internal gain $\Gamma g_\text{int}$ in Eq.~\eqref{eqn:internal_material_gain_02}.
The measured VSL data and the deduced gain values for $1.0$~wt.\% DCM in PMMA are summarized in Figs.~\ref{fig:gain_efficiency}(a) and \ref{fig:gain_efficiency}(b), respectively. A gain efficiency $\Gamma K$ of \SI{240}{\per\centi\meter\per\mega\watt} was determined. 

The gain efficiency $\Gamma K$ can also be estimated purely from material parameters. If concentration quenching is omitted, the gain efficiency can be approximated by\cite{Gozhyk.2015}
\begin{equation}
\label{eq:gain_efficiency}
K = \sigma_\text{em} \sigma_\text{abs} N \frac{\tau_\text{f}}{h \nu},
\end{equation}
where $h$ represents Planck's constant and $\nu$ the pump light frequency. 
In the following, we calculated the theoretical modal gain efficiency for DCM in PMMA, for which the required material parameter values are well studied and available in literature. The absorption and emission cross sections of DCM in PMMA at a wavelength of \SI{450}{\nano\meter} and \SI{600}{\nano\meter} are \SI{130e-18}{\square\centi\meter} and \SI{320e-18}{\square\centi\meter}, respectively \cite{Forget.2013}. The fluorescence decay time $\tau_\text{f}$ of DCM in PMMA is approximately \SI{2}{\nano\second} \cite{Gozhyk.2015, Green.2015}. The confinement factor $\Gamma$ can be readily derived from eigenmode calculations, for which we used the polymer refractive index and neglected the doping contribution, resulting in a confinement factor for the TE-like polarization of approximately $\Gamma=0.8$. In summary, a $1.0$~wt.\% DCM doped PMMA thin film results in a theoretical modal gain efficiency of 
% 0.8 * 130e-18 cm2 * 320e-18 cm2 * 2 ns * 0.01 * 1.19 g/cm2*6.023e23/303.36 g/Mol / E_photon
% E_photon = 6.626e-34J * 2.998e8m/s / 450nm
$\Gamma K=\SI{3500}{\per\centi\meter\per\mega\watt}$. 
These calculations significantly overestimate the gain efficiency compared to the experimentally determined value. 

If we apply Eq.~\eqref{eq:gain_efficiency_quenched} of our gain model, which takes concentration quenching into account, the gain efficiency amounts to 
% Calculation (Mathematica):
% w = 0.001;
% w0 = 3.4;
% rho = 1.19;
% avogadro = 6.023*10^23;
% molarMass = 303.36;
% eta0 = 0.33;
% beta13 = 2.0;
% tauRad = 7.1*10^-9;
% sigmaEmission = 130*10^-18;
% sigmaAbs = 320*10^-18;
% gamma = 0.8;
% EPhoton = 6.626*10^-34*2.998*10^8/(450*10^-9);
% In[27]:= w gamma rho avogadro/
%   molarMass eta0 (1 - 
%    Exp[-beta13 (w^(-1/3) - (w0)^(-1/
%           3))]) tauRad sigmaAbs sigmaEmission/EPhoton
% Out[27]= 0.000417337
$\Gamma K=\SI{417}{\per\centi\meter\per\mega\watt}$. In this calculation, we used the experimentally determined quenching coefficients for DCM doped PMMA summarized in Table \ref{tab:quenching_parameters}. The values of the radiative lifetime $\tau_\text{rad}=\SI{7.1}{\nano\second}$ and the quantum yield $\eta_0=0.33$ are taken from Green et al. \cite{Green.2015}. The proposed quenching model strongly reduces the discrepancy between theory and measured data. This indicates that the linear model for the dependence of the gain efficiency on the dye concentration is only valid for low dye concentrations at which concentration quenching is negligible. For increased dye concentrations and close to the material gain peak value, quenching phenomena need to be included in the description of the system. 

 %%%%%%%%%%%%%%%%%%%%%%%%%%  Conclusion  %%%%%%%%%%%%%%%%%%%%%%%%%%
\section{Conclusion and outlook}
In this paper we present a material gain model for organic dye-doped polymers that is able to describe concentration quenching. This model was applied to measurement results of the novel organic dye molecule PMN and the reference dye DCM. For both dyes concentration quenching was well described and model parameters have been extracted. Additionally, the presented gain model reduces the overestimation of the gain efficiency compared to a linear model, where intermolecular interaction is not included. Modal gain values were obtained by a refined VSL method. This method accounted for geometric effects arising from the narrow pump stripe and the numerical aperture of the light collecting fiber. Our results show that high power laser diodes can be used as pump light source for gain measurements in organic dye-doped polymer thin films using the VSL method. For PMN we reached a material internal gain $\Gamma g_\text{int}$ up to \SI{43}{\per\centi\meter} in PMMA at a dye concentration of $5.0$~wt.\%, whereas for DCM we reached \SI{34}{\per\centi\meter} at a dye concentration of $1.5$~wt.\%. Taking losses into account, modal gain of \SI{39}{\per\centi\meter} for PMN and \SI{16}{\per\centi\meter} for DCM were obtained. The high achievable gain and reduced concentration quenching compared to DCM make PMN a promising organic solid-state laser dye candidate. The influence of the polymer matrix on gain properties has been investigated for PMN. Even though modal gain was reduced to \SI{32}{\per\centi\meter} in PC, the higher refractive index could be an interesting material property for photonic integrated circuit designs. 

In a next step, the material gain model will be applied to a wider range of dyes and polymers. Moreover, future work will also address the influence of the pump intensity and the duration of the excitation pulse on concentration quenching. Furthermore, laser dye parameters such as photostability and threshold will be investigated. Ultimately, we envisage the development of a laser diode pumped integrated organic solid-state laser based on the novel dye molecule. 

%%%%%%%%%%%%%%%%%%%%%%%%%%  Funding  %%%%%%%%%%%%%%%%%%%%%%%%%%
\section*{Funding}
Austrian Research Promotion Agency (FFG) (850649).

\section*{Acknowledgments}
This research has received funding through the grant PASSION (No. 850649) from the Austrian Research Promotion Agency (FFG). The authors wish to thank \emph{ams~AG} for supplying high quality silicon substrates.

%%%%%%%%%%%%%%%%%%%%%%% References %%%%%%%%%%%%%%%%%%%%%%%%%

%%%%%%%%%% If using BibTeX:
%\bibliography{bibliography}

%%%%%%%%%% If preparing manually:

\end{document}